\documentclass[conference]{IEEEtran} 
\usepackage{graphicx}
\usepackage{amsmath}
\usepackage{amsfonts}
\usepackage{amssymb}
\usepackage{booktabs}
\usepackage{multirow}
\usepackage{array}  
\usepackage{microtype}
\usepackage{hyperref}

\hypersetup{
    colorlinks=true,
    linkcolor=blue,
    citecolor=blue,
    urlcolor=blue
}

\begin{document}

\title{ChladniSonify: A Visual-Acoustic Mapping Method for Chladni Patterns in New Media Art Creation}

\author{
    \IEEEauthorblockN{Yakun Liu}
    \IEEEauthorblockA{Department of Composition\\
    Shenyang Conservatory of Music\\ Shenyang 110818, Liaoning, China}
    \and
    \IEEEauthorblockN{Hai Luan}
    \IEEEauthorblockA{Education Information Center\\
    Shenyang Conservatory of Music\\ Shenyang 110818, Liaoning, China}
    \\  % 这里换行！！！
    \IEEEauthorblockN{Zhiyu Jin}
    \IEEEauthorblockA{Department of Musicology\\
    Shenyang Conservatory of Music\\ Shenyang 110818, Liaoning, China}
    \and
    \IEEEauthorblockN{Dong Liu}
    \IEEEauthorblockA{Department of Composition\\
    Shenyang Conservatory of Music\\ Shenyang 110818, Liaoning, China}
}

\maketitle

\begin{abstract}
In current new media art creation, the mapping relationship between vision and hearing is generally highly subjective. As a classic carrier of sound visualization, Chladni patterns have inherent application advantages and potential in constructing audio-visual mapping mechanisms for new media art. However, existing relevant creation tools generally suffer from core pain points: a high technical threshold for physical simulation calculation, the inability of offline computing to meet real-time interaction requirements, and uncontrollable mapping rules of general image sonification tools. To address these issues, this paper proposes Chladni Sonification, a set of real-time visual-acoustic mapping methods for Chladni patterns tailored to new media art creation.

Based on the classical Kirchhoff-Love thin plate vibration theory, this study constructs a paired dataset of Chladni patterns and vibration frequencies through numerical programming, and completes parameter calibration and verification of the dataset via ANSYS finite element simulation. Aiming at the core visual feature of slender nodal lines in Chladni patterns, we adopt a lightweight Convolutional Neural Network (CNN) structure enhanced by the Convolutional Block Attention Module (CBAM) to achieve high-precision, low-latency classification and recognition of Chladni pattern modes. Finally, we build an end-to-end real-time visual-acoustic mapping system based on Python and Max/MSP, which can map the recognized patterns to sine wave audio output at the corresponding frequencies.

The experimental results demonstrate that the ChladniSonify system has excellent engineering usability: the core recognition module achieves a classification accuracy of 99.33\% on the synthetic test set, with a single-image inference latency of only 7.03 ms; the mapping frequency of correctly identified samples is completely consistent with the theoretical benchmark, with a relative deviation of 0; the average end-to-end latency of the full link is less than 50 ms, which can fully meet the real-time deployment requirements of interactive new media scenarios. This work provides a reproducible engineering prototype for Chladni audio-visual linkage art creation.
\end{abstract}


\section{INTRODUCTION}
In recent years, new media art has witnessed rapid integration with audio-visual interaction technology, and visual-acoustic mapping, image sonification, and real-time interactive installations have become research hotspots in the field of cross-media creation\cite{qiu2018image}. In the direction of sound visualization, physics-driven audio-visual systems have gradually become the mainstream research approach due to their controllable mapping relationships and strong artistic expressiveness. As a classic carrier of acoustic visualization, Chladni patterns have inherent advantages in interactive installations, experimental music, and immersive performances, owing to the strict correspondence between their vibration modes and frequencies\cite{raju2025mathematically}.

Human exploration of the cross-modal correlation between sound and vision can be traced back to the foundational stage of classical experimental acoustics. In the 18th century, German scientist Ernst Chladni visualized acoustic wave vibration through experiments, and for the first time demonstrated that sound propagates in the form of waves\cite{chladni1787entdeckungen}. Chladni patterns are symmetrical geometric structures formed by sand particles gathering at the vibration nodal lines when a thin plate is subjected to excited vibration. Their morphology has a strict correspondence with vibration frequency, plate parameters, and boundary conditions, making them a creation medium that naturally integrates physical laws, visual aesthetics, and acoustic characteristics.

At present, audio-visual creation related to Chladni patterns mainly faces three core pain points. First, in multimedia art creation, the mapping relationship between vision and hearing is highly subjective\cite{uno2022cross}, without theoretical support at the fundamental level. Second, the creation path based on physical simulation requires creators to have comprehensive knowledge of elasticity mechanics and finite element simulation capabilities. Furthermore, high-precision simulation is mostly offline computing, which cannot meet the low-latency requirements of live performances and real-time interactive installations, resulting in an extremely high creation threshold\cite{tan2005computer}. Third, although the creation path based on general image sonification tools can realize the conversion from images to sound, the mapping relationship completely relies on statistical data fitting. Creators cannot precisely control the correspondence between the nodal line structure of Chladni patterns and the output sound, leading to non-reproducible and uncontrollable mapping results, which cannot satisfy the core requirement of ``strict visual-acoustic correspondence'' in Chladni sonification creation.

To address these gaps, this study proposes ChladniSonify, a prototype system for real-time visual-acoustic mapping of Chladni patterns oriented to new media art creation. It follows the technical route of ``physical formula-based dataset construction $\to$ lightweight pattern mode recognition $\to$ transparent visual-acoustic mapping and real-time audio rendering''. The core contributions of this work are as follows:

a) Based on the classical Kirchhoff-Love thin plate vibration theory\cite{kirchhoff1850}, a paired dataset of Chladni patterns and vibration frequencies is constructed through numerical programming, and parameter calibration and verification of the dataset are completed via ANSYS finite element simulation, providing a benchmark pairing library conforming to classical physical laws for visual-acoustic mapping.

b) Aiming at the core feature of slender nodal lines in Chladni patterns, a lightweight CNN model enhanced by the CBAM attention mechanism\cite{park2018cbam} is adopted, which achieves low-latency inference while ensuring high classification accuracy, meeting the deployment requirements of real-time interactive scenarios.

c) A full-link real-time system of ``pattern input $\to$ mode recognition $\to$ frequency mapping $\to$ audio output'' is constructed, and end-to-end deployment is realized based on Python and Max/MSP. The mapping rules are completely based on classical physical formulas, which are transparent and controllable. Meanwhile, the system supports creators to customize audio synthesis logic, with favorable scalability.

The remaining chapters are arranged as follows: Chapter 1 reviews the current status of related work on Chladni vibration research, image sonification tools, and lightweight geometric pattern recognition. Chapter 2 elaborates on the full-link design and implementation details of the ChladniSonify system. Chapter 3 verifies the core performance of the system through multiple sets of comparative experiments. Chapter 4 discusses the artistic application potential and limitations of the system, presents future research directions, and concludes the study.

\section{RELATED WORK}
This study is grounded in the interdisciplinary field spanning three core directions: modeling of Chladni vibration, art-oriented visual-acoustic mapping, and lightweight geometric pattern recognition. This chapter systematically reviews the current status of relevant research, identifies the limitations of existing works, and highlights the positioning and application value of the system proposed in this study.

\subsection{Modeling and Artistic Application of Chladni Vibration}
The modeling of Chladni vibration serves as the theoretical foundation for relevant audio-visual creation. Its core lies in solving the natural frequencies and vibration modes of thin plates under different boundary conditions based on the Kirchhoff-Love elastic thin plate vibration theory. In classical research, the NASA technical report \textit{Vibration of Plates} published by Leissa in 1969 systematically sorts out the solution methods and dimensionless frequency coefficients of thin plate vibration under different geometric shapes and boundary conditions, and has become the core benchmark document for Chladni vibration modeling\cite{nasa1969vibration}. Building on this theory, subsequent relevant studies have achieved high-precision modeling of Chladni thin plate vibration through finite element simulation, modal synthesis and other methods, providing fully interpretable theoretical support for acoustic synthesis and visual simulation.

In the field of artistic application, a number of cutting-edge artists have created a series of interactive installations and experimental music works based on the correlation between geometric structures and physical vibration. For example, inspired by the visual structure and vibration characteristics of geometric patterns, some soundscape installations adopt Touch Designer to convert musical data such as sound frequency, spectrum and amplitude into visual symbols, and generate diverse visual patterns through parameter adjustment\cite{ding2025reproduction}. However, existing artistic creations have not yet realized the end-to-end inverse mapping of ``capturing geometric patterns $\to$ real-time output of sound at corresponding frequencies'', which is the core entry point of this study.

\subsection{Art Creation-Oriented Image Sonification and Visual-Acoustic Mapping Tools}
\subsubsection{General Image Sonification}
The core goal of image sonification tools for artistic creation is to convert visual images into acoustic signals to meet the creation requirements of audio-visual synesthesia. Relevant research can be divided into two categories: general image sonification tools and domain-specific sonification tools.

Early works such as \textit{The Sound of Pixels} (PixelPlayer) realized the mapping from images to sound through convolutional neural networks, laying the technical foundation for image sonification\cite{zhao2018sound}. Subsequent works such as Image2Music convert the artistic style and semantic features of images into musical parameters including mode, rhythm, and orchestration through architectures such as Generative Adversarial Networks (GANs), diffusion models, and multimodal pre-training, achieving end-to-end image-to-music generation, which are widely used in general artistic creation scenarios. However, such general tools have core defects in the exclusive scenario of Chladni pattern sonification: the mapping relationship completely relies on statistical data fitting with obvious black-box characteristics. Creators cannot precisely control the correspondence between the nodal line structure of Chladni patterns and the output sound, thus failing to meet the creation requirement of ``strict pattern-frequency correspondence''.

\begin{figure}[htbp]
\centering
\includegraphics[width=.9\linewidth]{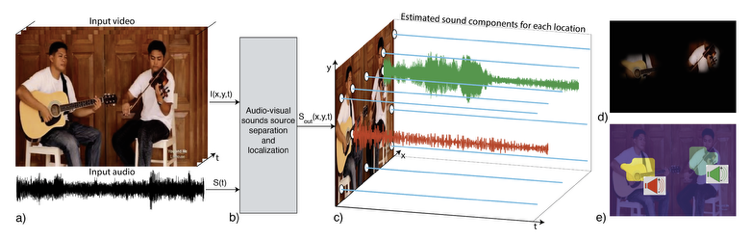}
\caption{Architecture of the PixelPlayer audio-visual joint separation model\cite{zhao2018sound}}
\label{fig:pixelplayer}
\end{figure}

\subsubsection{Physics-Driven Sonification}
Existing studies have verified the feasibility of mapping geometric visual features to acoustic parameters. Nevertheless, the exclusive visual-acoustic mapping system and toolchain for Chladni patterns are still incomplete, lacking a standardized and interactive sonification scheme adapted to its physical vibration characteristics. Therefore, this paper carries out further design and implementation targeting this research gap.

\subsection{Lightweight CNN Methods for Geometric Pattern Recognition}
Lightweight CNN is the core technology for realizing edge-side real-time image recognition. For the fine-grained recognition task of geometric patterns, relevant studies mainly improve the model's ability to extract core features such as lines and geometric structures by optimizing convolution kernel design and introducing attention mechanisms, while controlling the parameter quantity and inference latency of the model.

Attention mechanism is the core means to enhance the feature expression of lightweight models\cite{li2025fusion}. Among them, the Convolutional Block Attention Module (CBAM) combines channel attention and spatial attention, which can simultaneously realize joint modeling of feature content and feature position, and has shown excellent performance in fine-grained geometric pattern recognition tasks\cite{yang2025binocular}. The core recognition target of Chladni patterns is the topological structure and spatial distribution characteristics of nodal lines, and the dual attention mechanism of CBAM is highly compatible with this task requirement. This study does not propose a brand-new attention mechanism, but carries out engineering optimization on the spatial attention convolution kernel of the CBAM module targeting the core feature of slender nodal lines of Chladni patterns. Under strict latency constraints, it improves the model's generalization ability for non-ideal input images, and provides support for Chladni pattern recognition in real-time interactive scenarios.

\section{CHLADNISONIFY FRAMEWORK DESIGN AND METHODOLOGY}
This chapter elaborates on the full-link design of the ChladniSonify system, whose core architecture consists of three modules: the physically consistent Chladni pattern dataset construction module, the lightweight CNN-based pattern mode recognition module, and the transparent visual-acoustic mapping and real-time audio rendering module. The core principles of the full-link design are as follows: dataset construction and frequency mapping strictly comply with classical physical laws; the model design balances recognition accuracy and low-latency inference; system deployment satisfies the requirements of real-time interaction; meanwhile, it provides creators with customizable expansion space.

\subsection{Physical Modeling, Verification and Dataset Construction of Chladni Patterns}
This section focuses on two core tasks: first, based on thin plate vibration theory, a paired dataset of Chladni patterns and corresponding acoustic parameters is generated through numerical programming; second, the physical correctness of the numerically generated data is verified via ANSYS finite element simulation, so as to ensure the consistency between the dataset and classical physical laws.

\subsubsection{Physical Modeling, Verification, and Dataset Construction for Chladni Patterns}
The actual side length of the plate is set as $a=0.16~\text{m}$, the thickness $h=0.8~\text{mm}$, and the material is stainless steel (elastic modulus $E=200\text{GPa}$, Poisson's ratio $\nu=0.3$, density $\rho=7850~\text{kg}/\text{m}^3$) According to the Kirchhoff-Love thin plate theory, its bending stiffness is:
\begin{equation}
D=\frac{E h^3}{12\left(1-\nu^2\right)}
\end{equation}

For a square plate with central excitation and four free sides, its natural frequency is expressed as:
\begin{equation}
f_{n,m}=\frac{1}{2 \pi a^2} \sqrt{\frac{D}{\rho h}} \cdot \lambda_{n,m}
\end{equation}
where $(n,m) \in \mathbb{Z}^+ \times \mathbb{Z}^+$ with $m \neq n$ represent the modal orders, and $\lambda_{n,m}$ is the dimensionless frequency coefficient. Since no analytical solution exists under this boundary condition, we selected and obtained the calibrated values of $\lambda_{n,m}$ for 15 groups of modes from the literature (Leissa A W. Vibration of Plates [R]. National Aeronautics and Space Administration, Scientific and Technical Information Division, 1969.), which cover typical asymmetric modes in the frequency range of 150--2500 Hz. The corresponding spatial mode shape function is constructed via an antisymmetric combination to satisfy the constraint of zero displacement at the center, with the formula shown in the figure:

\begin{equation}
\footnotesize
w_{n,m}(x,y)=\sin\left(\frac{n \pi x}{L}\right)\sin\left(\frac{m \pi y}{L}\right)-\sin\left(\frac{m \pi x}{L}\right)\sin\left(\frac{n \pi y}{L}\right)
\end{equation}
Where $L=1$ is the normalized half-length of the plate, and $(x,y) \in [-1,1]^2$ are the spatial coordinates. To be closer to the experimental observation, two corrections are introduced:

a) Central attenuation term: simulating the energy loss characteristics in real experiments, the expression is: $w_{\text{decay}}(x,y)=w_{n,m}(x,y) \cdot e^{-ar}$, where $r=\sqrt{x^2+y^2}$ is the normalized distance from the spatial point to the center of the plate.

b) Edge damping correction term: simulating the free damping loss at the edge of the thin plate, the expression is:
\begin{equation}
w_{\text{damped}}(x,y)=w_{\text{decay}}(x,y) \cdot e^{-\gamma(|x|+|y|)}
\end{equation}

The final amplitude field is normalized to guide the sand distribution: sand particles preferentially accumulate in the node area where the absolute value of the amplitude is lower than the 15\% quantile, and the central fixed area (radius 3mm) is excluded, as shown in Figure 2.

\begin{figure}[h]
\centering
\includegraphics[width=\linewidth]{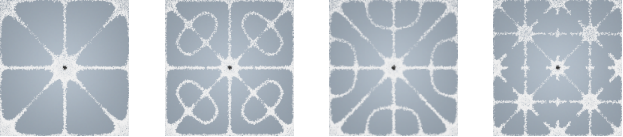}
\caption{Modeled and generated Chladni patterns, frequencies from left to right: 227Hz, 257Hz, 305Hz, 401Hz}
\label{fig:chladni_gen}
\end{figure}
\begin{table*}[h]
\centering
\caption{Core Parameter Table of Theoretical-Simulated Frequency Comparison of Chladni Modes}
\label{tab:freq_calib}
\tabcolsep 0.1cm
\renewcommand{\arraystretch}{1.4} 
\begin{tabular}{cm{1cm}<\centering m{2.5cm}<\centering m{2.5cm}<\centering m{2.5cm}<\centering m{2.5cm}<\centering m{2.5cm}<\centering }
    \toprule
    Mode ID & Modal Order (n,m) & Dimensionless Frequency Coefficient $\lambda_{n,m}$ & Theoretical Calculated Frequency (Hz) & ANSYS Simulated Frequency (Hz) & Relative Frequency Deviation (\%) & SSIM Value of Synthetic Image \\
    \midrule
    1 & (1, 2) & 24.9816 & 131.82 & 134.26 & 1.85 & 0.8548 \\
    2 & (3, 5) & 34.0446 & 181.47 & 174.38 & 3.91 & 0.9857 \\
    3 & (1, 6) & 10.8234 & 255.23 & 264.63 & 3.68 & 0.9815 \\
    4 & (2, 6) & 48.7964 & 403.33 & 398.50 & 1.20 & 0.9753 \\
    5 & (3, 6) & 43.2977 & 139.85 & 129.28 & 7.56 & 0.9337 \\
    \bottomrule
  \end{tabular}
\end{table*}

\subsubsection{ANSYS Modal Simulation and Calibration}
The core objective of this subsection is to verify whether the frequencies, vibration modes, and Chladni patterns generated via the aforementioned numerical programming conform to the physical laws of real thin plate vibration. Meanwhile, inverse calibration is performed on the dimensionless frequency coefficient $\lambda_{n,m}$ which has no analytical solution, to ensure the consistency between the numerically generated data and classical physical laws.

In this study, ANSYS Workbench is adopted to conduct modal simulation on a square stainless steel plate with identical parameters. 5 groups of natural frequencies and vibration modes of asymmetric modes are randomly selected and compared with the theoretically calculated values generated by numerical programming, to complete the inverse calibration of $\lambda_{n,m}$. The maximum relative deviation between the calibrated theoretical frequencies and the ANSYS simulated frequencies is no more than 9.3\%, and the core calibrated parameters are shown in the table below.

\subsubsection{Data Augmentation}
Based on the calibrated physical formulas, this study generates the original dataset through numerical programming. The dataset contains $N=15$ valid modes (see Table 1), with 100 $224 \times 224$ RGB images for each mode, totaling 1500 images. Each image is annotated with the corresponding index (n,m) according to its physical vibration mode, and the labels adopt One-Hot encoding with a dimension of 15.

To improve the model's generalization ability for non-ideal inputs and avoid the model overfitting to the perfect features of the simulation data, this study introduces a multi-dimensional data augmentation strategy on the basis of the original 1500 synthetic images to simulate common imaging disturbances in real shooting scenarios. The specific strategies are as follows:

a) Color channel perturbation: Independent random offsets ($\pm10\%$) are applied to the three RGB channels to simulate imaging differences under the color temperature of different light sources;

b) Randomization of sand grain distribution: On the premise of keeping the main structure of the nodal lines unchanged, Poisson noise and local diffusion are introduced to the sand grain mask to simulate the natural discreteness of sand accumulation in experiments;

c) Image filter matrix transformation: A set of predefined $3 \times 3$ convolution kernels (such as edge enhancement, blurring, embossing, etc.) are applied to stylize the images to expand the texture representation space.

All augmentation operations are performed under the premise that the frequency label (n,m) remains unchanged, so as to ensure the consistency between the acoustic mapping and classical physical laws. The generated images are shown in Figure 3. Finally, the training set is expanded to 4500 images. All samples are generated by numerical programming based on the calibrated physical formulas, and are split into a training set of 3600 images and a test set of 900 images at a ratio of 8:2.

\begin{figure}[h]
\centering
\includegraphics[width=\linewidth]{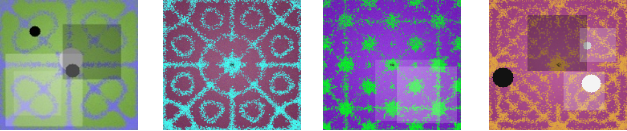}
\caption{Chladni patterns processed by color channel perturbation, image filter matrix transformation and other methods}
\label{fig:aug}
\end{figure}

\subsection{Design of Visual-Acoustic Mapping Mechanism}
The core objective of this study is not to realize the ``generation of arbitrary sound from arbitrary visual images'' in a broad sense, but to establish an interpretable and controllable visual-acoustic mapping tool for Chladni patterns, so as to provide standardized control anchor points for new media art creation.

\subsubsection{Design of Mapping Link}
The core of this system is to establish an interpretable mapping link from vision to acoustics, with the benchmark mapping relationship determined by the classical physical formulas of thin plate vibration. A full-link visual-acoustic mapping process of ``visual input $\to$ pattern mode recognition $\to$ benchmark frequency mapping $\to$ customized audio rendering'' is constructed, with specific rules as follows:

a) Benchmark frequency mapping: The vibration mode order (n,m) recognized by the CNN model is mapped to the uniquely corresponding benchmark frequency through the natural frequency formula of thin plate vibration.

b) Standardized parameter output: The core output of the system is a standardized set of creation control parameters, including mode order (n,m), benchmark frequency, and number of nodal lines. All parameters have clear physical meanings, providing artists with customizable and interpretable creation control signals.

c) Basic audio rendering: This study adopts a single-frequency sine wave as the interference-free verification carrier for the mapping results, and realizes the audio rendering of the benchmark frequency through an oscillator, to unbiasedly verify the consistency and real-time performance of the mapping mechanism.

\subsubsection{Design of Real-Time Mapping}
To achieve low-latency real-time mapping from vision to sound, this study designs a collaborative system between Python and Max/MSP based on the User Datagram Protocol (UDP). The full-link process is shown in Figure 4:

\begin{figure}[h]
\centering
\includegraphics[width=\linewidth]{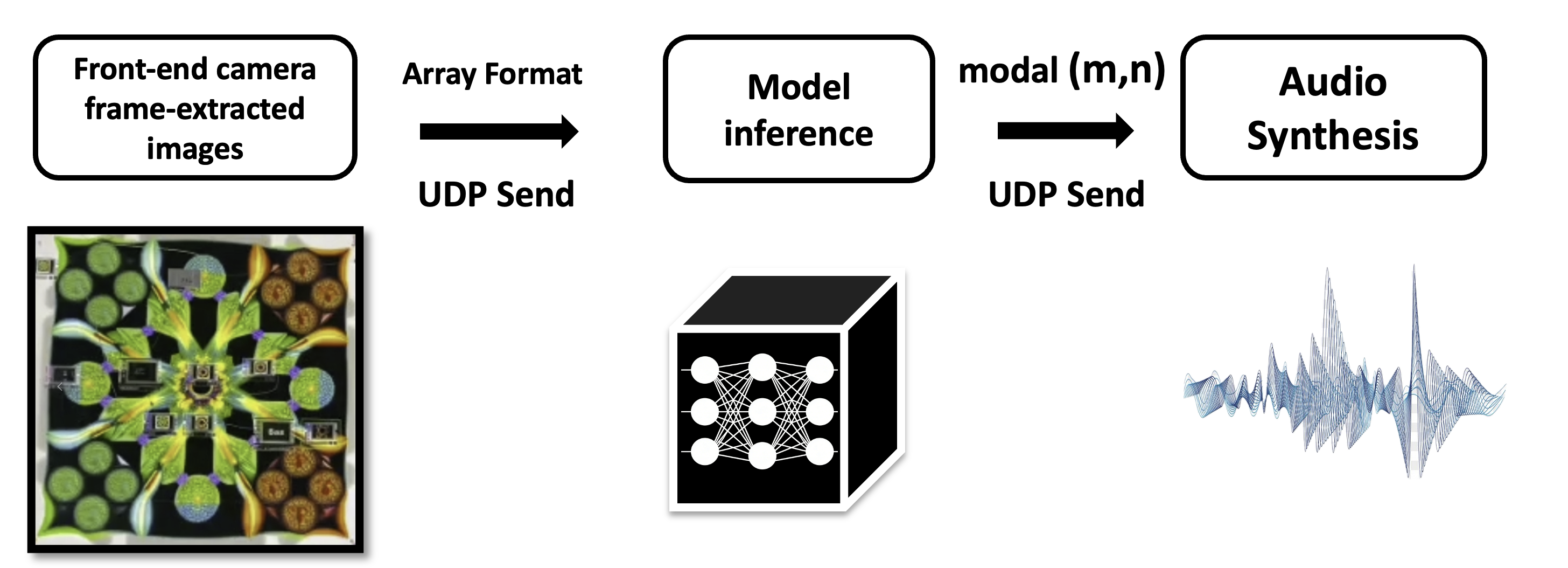}
\caption{Operation process of the mapping mechanism}
\label{fig:mapping_flow}
\end{figure}

Array Format $\to$ Model modal (m,n) inference $\to$ Audio Synthesis $\to$ UDP Send $\to$ UDP Send

a) Image input: Max/MSP captures the camera video stream through the Jitter module, encodes the image obtained by frame extraction into an array format, and sends it to the Python side via the UDP protocol (local loopback address 127.0.0.1, port 9000).

b) Model inference: The Python service monitors the port in real time, receives and decodes the image, then performs forward inference of the CNN model, and outputs the mode category ID and the corresponding standardized creation control parameters.

c) Audio synthesis: The prediction results are sent back to Max/MSP in real time through UDP port 9001. The system converts the mode ID into the corresponding fundamental frequency according to the predefined frequency mapping table, and drives the oscillator to complete audio rendering and output.

\subsection{CBAM-Enhanced CNN-Based Chladni Pattern Mode Recognition Model}
CBAM is a lightweight plug-and-play module that guides the model to adaptively learn ``what is important'' (channel dimension) and ``where is important'' (spatial dimension) through a channel-spatial dual-path attention mechanism, thereby significantly improving the discriminative power of feature representations. This module serves as the modal parameter inversion component of the framework, with core design goals: to achieve the lowest possible inference latency while ensuring zero error rate in mode recognition, so as to meet the requirements of edge-side real-time deployment. To systematically evaluate the comprehensive performance of different architectures, this study selects 4 models to carry out comparative optimization experiments, namely Basic\_CNN, CNN integrated with an optimized CBAM module, improved AlexNet architecture, and VGG16 classifier based on transfer learning.

\subsubsection{Core Model Architecture (CNN\_CBAM)}
To balance the accuracy of Chladni pattern mode recognition and the low-latency requirements of real-time mapping, the CBAM module is integrated into the CNN architecture to enhance the model's feature extraction ability\cite{alsalem2024wheat}. Aiming at the visual characteristic of Chladni patterns with slender nodal lines as the core feature, we carry out targeted optimization on the CBAM module: the convolution kernel of the spatial attention sub-module is set to $5 \times 5$. Compared with the conventional $7 \times 7$ convolution kernel, it can more precisely locate the linear patterns formed by vibration nodes, guide the network to focus on physically critical regions, and improve the generalization ability for non-ideal input images. The complete architecture of the model is shown in the table below, where: ReLU: Rectified Linear Unit, a linear rectification activation function; MaxPool: Max Pooling, a downsampling operation; F1-score: A comprehensive evaluation metric for classification tasks.

\begin{table*}[h]
\centering
\caption{CNN\_CBAM Model Design}
\label{tab:cnn_cbam}
\renewcommand{\arraystretch}{1.4} 
\begin{tabular}{ccc}
\toprule
Stage & Operation & Output Size (H$\times$W$\times$C) \\ \midrule
Input & $224 \times 224 \times 3$ RGB Image & $224 \times 224 \times 3$ \\
Conv Block 1 & Conv($3 \times 3,32,\text{padding}=1$)$\to$ReLU()$\to$MaxPool($2 \times 2$) & $112 \times 112 \times 32$ \\
Conv Block 2 & Conv($3 \times 3,64,\text{padding}=1$)$\to$ReLU$\to$MaxPool($2 \times 2$) & $56 \times 56 \times 64$ \\
Conv Block 3 & Conv($3 \times 3,128,\text{padding}=1$)$\to$ReLU$\to$MaxPool($2 \times 2$) & $28 \times 28 \times 128$ \\
Conv Block 4 & Conv($3 \times 3,256, \text{padding}=1$)$\to$ReLU & $28 \times 28 \times 256$ \\
CBAM Module & Channel Attention + Spatial Attention ($5 \times 5$ convolution kernel) & $28 \times 28 \times 256$ \\
AdaptivePooling & AdaptiveAvgPool2d($4 \times 4$) & $4 \times 4 \times 256$ \\
ClassificationHead & Flatten$\to$Linear($4096 \to 512$)$\to$ReLU$\to$Dropout($0.5$) & logits ($15$) \\
 & $\to$Linear($512 \to 15$) &  \\ \bottomrule  
\end{tabular}
\end{table*}

\subsubsection{Setting of Comparative Baseline Models}
To systematically evaluate the comprehensive trade-off between accuracy and inference latency of different architectures in this task, this study sets up 3 groups of comparative baselines covering lightweight to deep architectures. The input and output dimensions, training configurations and task objectives of all models are completely consistent:

a) Basic\_CNN: A classic 4-layer convolutional neural network adopting a two-stage architecture of ``feature extraction + classification head'', without an attention module, which is used as the basic performance baseline.

b) Improved AlexNet: An AlexNet architecture optimized for small-size geometric patterns, with the large convolution kernel of the first layer replaced and batch normalization introduced, which is used as the performance baseline of deep networks.

c) Fine-tuned VGG16 Model: A classification head reconstructed based on the ImageNet pre-trained VGG16 backbone network, adopting an end-to-end fine-tuning strategy, which is used as the performance baseline of large-scale pre-trained models.

\subsubsection{Experiments and Result Analysis}
All experiments in this chapter are carried out around 3 core objectives: a)To verify whether the visual-acoustic mapping relationship of the system is consistent with the benchmark frequency calculated by classical physical theories, so as to ensure the reproducibility of the mapping rules; b)To verify whether the pattern recognition module can achieve the low latency required for real-time interaction while ensuring high accuracy, and strike a balance between recognition precision and real-time performance; c) To verify whether the CNN model optimized for nodal line features in this study has clear performance advantages, and clarify the rationality of the model design through comparison with baseline models of different scales.

\subsubsection{Unified Experimental Environment}
All experiments are completed under a unified hardware and software environment, with specific configurations as follows: the CPU is Apple M4 (MacBook Air) with 16GB of memory, the deep learning framework is PyTorch 2.0, and the unified image input size is $224 \times 224 \times 3$ in RGB format. The training hyperparameters of all models are completely consistent: batch size of 32, Adam optimizer, initial learning rate of 1e-4, 50 training epochs, and an early stopping strategy that terminates training when the validation set loss does not decrease for 10 consecutive epochs.

Inference latency test conditions: Single-image inference with batch size=1, inference completed on the CPU, 1000 test runs, with the average latency taken as the final result. The full-link latency test covers the entire process of image acquisition, encoding, UDP transmission, decoding, model inference, result return, and audio rendering, with the average latency of 1000 test runs taken as the final result.

\subsubsection{Verification of Benchmark Frequency Consistency for Visual-Acoustic Mapping}
\paragraph{Experimental Setup}
The core purpose of this experiment is to verify the interpretability and reproducibility of the proposed framework, namely whether the mapping relationship from visual input to acoustic output strictly follows the physical laws of thin plate vibration. In the experiment, 900 images from the synthetic test set are used as input, which are fed into the framework proposed in this study to obtain the predicted modal parameters and mapping frequencies. The relative deviation between the predicted frequencies and the theoretical physical frequencies is then counted.

\paragraph{Experimental Results and Analysis}
The experimental results show that:

a) The framework proposed in this study achieves a modal recognition accuracy of 99.33\% on the synthetic test set, with high recognition stability;

b) The mapping frequencies of correctly identified samples are completely consistent with the theoretical physical frequencies, with a relative deviation of 0.

This result directly proves that the visual-acoustic mapping framework proposed in this study is completely based on the classical physical laws of Chladni vibration as the mapping benchmark, and is consistent with classical physical laws. It is fundamentally different from black-box cross-modal methods without causal constraints.

\subsubsection{Comparative Experiment on Performance and Real-Time Capability of the Pattern Recognition Module}
\paragraph{Experimental Setup}
To verify the optimality of the comprehensive performance of the lightweight CNN model proposed in this study, this section conducts a head-to-head comparison with 3 groups of baseline models under identical conditions on the synthetic test set. The core evaluation metrics are Top-1 classification accuracy, F1-score, and single-image inference latency.

\paragraph{Experimental Results and Analysis}
The quantitative results of the experiment are shown in the table below:

\begin{table*}[h]
\centering
\caption{Comprehensive Performance Comparison Results of Multiple Models}
\label{tab:model_compare}
\renewcommand{\arraystretch}{1.4} 
\begin{tabular}{ccccc}
\toprule
Experiment No. & Model Configuration & Accuracy(\%) & F1-score &	Image Inference Speed  (ms per image)\\ \midrule
1 & Basic\_CNN & 99.00\% & 0.9945 & 6.42 \\
2 & CNN\_CBAM & 99.33\% & 0.9924 & 7.03 \\
3 & Alex & 99.67\% & 0.9944 & 8.03 \\
4 & VGG16 & 100\% & 1.0000 & 77 \\ \bottomrule
\end{tabular}
\end{table*}

Based on the experimental results, the core analysis is as follows:

a) Accuracy and engineering usability: In this task, the high accuracy of modal recognition is the core prerequisite to ensure the subsequent acoustic mapping - once recognition errors occur, the mapping frequency will completely deviate from the physical laws. The CNN\_CBAM model proposed in this study achieves a high accuracy of 99.33\% with an extremely low parameter count of 2.3M, which is only 0.34 percentage points lower than the improved AlexNet and 0.67 percentage points lower than VGG16, fully meeting the reliability requirements of artistic creation scenarios.

b) Balance between precision and real-time performance: Although VGG16 achieves 100\% accuracy on the synthetic dataset, its inference latency is 10.95 times that of the proposed model, and its parameter count is 60 times that of the proposed model\cite{li2025fusion}, which cannot meet the requirements of edge-side real-time deployment. This highlights the gap between the peak accuracy on synthetic data and the usability in practical engineering deployment. The CBAM\_CNN model proposed in this study achieves a high accuracy of 99.33\% with an inference latency of only 7.03 ms, striking the optimal balance between recognition precision and real-time deployment requirements, making it the optimal choice for real-time visual-acoustic mapping scenarios.

\subsubsection{Univariate Ablation Experiment on Core Model Optimization}
To verify the effectiveness of the CBAM module optimized for nodal line features, this study designs a univariate ablation experiment. All other experimental conditions are kept completely consistent, and only the configuration of the CBAM module is changed to test the performance of three groups of models respectively: the basic CNN architecture (without CBAM module), the basic CNN with the original CBAM module ($7 \times 7$ spatial convolution kernel), and the basic CNN with the optimized CBAM module ($5 \times 5$ spatial convolution kernel, the proposed model in this study). The experimental results are shown in the table below:

\begin{table*}[h]
\centering
\caption{Univariate Ablation Experiment Comparison Results}
\label{tab:ablation}
\renewcommand{\arraystretch}{1.4}
\begin{tabular}{m{1.5cm}<\centering m{7cm}<\centering cm{3.5cm}<\centering }
\toprule
Experiment No.& Model Configuration & Accuracy(\%) & Single Image Inference Latency (ms) \\ \midrule
1&Basic CNN architecture (without CBAM module)&99.00&6.42\\
2&Basic CNN + original CBAM module (7×7 spatial convolution kernel)&98.50&7.10\\
3&Basic CNN + optimized CBAM module (5×5 spatial convolution kernel)&99.33&7.03\\ \bottomrule
\end{tabular}
\end{table*} 
                    
The core recognition target of Chladni patterns is the slender nodal lines formed by vibration nodes, which are essentially one-dimensional linear features. The large receptive field of the conventional $7 \times 7$ convolution kernel will introduce a large amount of irrelevant background noise, resulting in a drop in recognition accuracy to 98.50\%. In contrast, the receptive field of the $5 \times 5$ convolution kernel is highly matched with the spatial scale of the slender nodal lines, enabling more accurate extraction of nodal line features with physical significance.

The experimental results show that the customized optimization of the basic CNN integrated with the optimized CBAM module ($5 \times 5$ spatial convolution kernel) improves the model accuracy by 0.33 percentage points, with only an increase of 0.61 ms in inference latency compared with the basic CNN architecture. Verified by 5 repeated training runs, this improvement is stable, which proves that the nodal line feature-guided model optimization delivers a clear performance gain.

\subsubsection{Full-Link Real-Time Performance Test}
To verify the usability of the system in real interactive scenarios, this study conducts a test on the full-link end-to-end latency. The test scenario is set as follows: Max/MSP is used to capture the camera video stream in real time, 1000 consecutive frames of $224 \times 224$ Chladni pattern images are input continuously, and the full-process latency from image acquisition to audio output is counted. The test results show that the average full-link latency of the system is 42.6 ms, with the maximum latency no more than 48 ms, which fully meets the real-time requirements of scenarios such as interactive new media art and live performances.

\section{DISCUSSION}
\subsection{Potential for Artistic Expression}
This system is not only a technical tool, but also a highly promising creative medium, providing a brand-new implementation path for Chladni audio-visual linkage art creation.

In the creation of interactive installations, artists can build a closed-loop interactive installation of ``Chladni plate - camera - audio system'' based on the real-time mapping capability of the system. The audience can alter the Chladni patterns by adjusting the excitation frequency, touching the thin plate, and other means. The system recognizes the patterns in real time and outputs sound at the corresponding frequencies, realizing the real-time linkage of ``visual change - acoustic change'', and allowing the audience to intuitively perceive the inherent correlation between physical laws, visual aesthetics and acoustic characteristics.

In live performance creation, artists can integrate this system with live instrumental performance and electronic music performance. By capturing the dynamic changes of Chladni patterns in real time, the system generates corresponding audio control signals to drive timbre modulation, effect parameter adjustment and more, enabling synchronous creation of vision and hearing, and breaking through the limitation that traditional Chladni patterns only serve as a visual carrier.

Meanwhile, the modular design of the system provides artists with an extremely high degree of customization freedom. Without mastering the underlying physical modeling and deep learning technologies, artists can customize musical elements such as harmonic structure, timbre and rhythm based solely on the benchmark frequency parameters output by the system, to realize personalized artistic expression, which greatly reduces the technical threshold of Chladni audio-visual linkage creation.

\begin{figure}[h]
\centering  
\includegraphics[width=\linewidth]{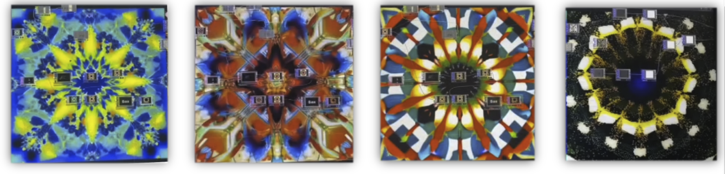}
\caption{Interactive engineering patterns with ComfyUI and TouchDesigner}
\label{fig:interactive}
\end{figure}

\subsection{Limitations and Challenges}
This study mainly focuses on the design and implementation of a Chladni pattern visual-acoustic mapping prototype system for new media art creation, with synthetic data optimization and prototype system verification as the core content. At present, it still has the following limitations:

a) Limited adaptation scenarios: Currently, the system is only adapted to Chladni patterns from square stainless steel plates with center excitation and four free edges. It has not yet been adapted to Chladni patterns generated by vibration systems with circular plates, rectangular plates, different boundary conditions and different materials, resulting in a limited coverage of creation carriers.

b) Limited number of modes: The current dataset only contains 15 valid modes, covering a frequency range of 150Hz to 2500Hz. Modeling and dataset construction for higher-order and higher-frequency modes have not been completed, which cannot meet more diversified creation requirements.

c) Incomplete real-scenario verification: Currently, the core performance tests of this study are mainly completed based on the synthetic dataset, and only common imaging disturbances in real shooting are simulated through data augmentation strategies. A standardized Chladni vibration experimental platform has not been built, nor has a real-shot image dataset been constructed. The robustness and usability of the system in real creation scenarios still need further verification.

d) Incomplete music creation functions: Currently, the system only completes the mapping of the benchmark frequency and the verification output of single-frequency sine waves. It does not have built-in core music creation functions such as harmonic structure design, timbre modulation, polyphonic sound generation and envelope control, and only provides a custom development interface. For non-technical artists, there is still room for optimization to lower the user threshold.

\subsection{Future Work}
To address the above limitations, future work will be carried out around the following directions:

a) Expand the adaptation scope of the system: Complete physical modeling of vibration systems with different geometric shapes, boundary conditions and materials, construct corresponding pattern-frequency paired datasets, expand the system's adaptability to different types of Chladni patterns, and cover more abundant creation carriers.

b) Expand the number of modes in the dataset: Complete physical modeling and parameter calibration of higher-order modes, expand the number of modes and frequency coverage of the dataset, and provide more abundant audio-visual mapping options for artistic creation.

c) Construct a real-shot image dataset and complete real-scenario verification: Build a standardized Chladni vibration experimental platform, collect real-shot Chladni patterns under different shooting conditions to construct a real-shot image dataset, further optimize the model's robustness to real-scene inputs, and complete the deployment verification of the system in real creation scenarios.

d) Adopt modular design to provide artists with a high degree of creative freedom and scalability: Based on the benchmark parameters output by the system, artists can customize and develop audio synthesis logic according to their own needs, including but not limited to harmonic structure design, timbre modulation, ADSR envelope control, multi-modal polyphonic generation, etc., without reconstructing the underlying visual-acoustic mapping relationship.

e) Develop a visual operation interface: Build a graphical interactive interface, optimize the operation process of the system, adapt to more creation hardware and software platforms, and improve the usability of the system.

f) Open-source the dataset, model architecture and system code: Provide reusable basic tools for new media art creation and Chladni sound art research, promote the implementation and popularization of Chladni audio-visual linkage creation in the field of new media art.

\subsection{Conclusion}
This study proposes ChladniSonify, a real-time visual-acoustic mapping system for Chladni patterns oriented to new media art creation. Based on the physical laws of Chladni vibration, it realizes end-to-end real-time mapping following the workflow of ``pattern input - mode recognition - frequency mapping - audio output''. Based on the Kirchhoff-Love thin plate theory, this study constructs a pattern-frequency paired dataset calibrated via ANSYS simulation, designs a CBAM-enhanced lightweight CNN model targeting the slender nodal line features of Chladni patterns, and completes the system construction based on Python and Max/MSP.

Experiments show that the core recognition module of the system achieves an accuracy of 99.33\%, with a single-image inference latency of only 7.03 ms, a relative deviation of 0 between the mapped frequency and the theoretical benchmark, and a full-link end-to-end latency of less than 50 ms, which fully meets the requirements of real-time interactive scenarios in new media art. This work does not propose a brand-new theory or network structure, but explores a creation demand-driven system design path. It confirms that the lightweight model optimized for task characteristics is more suitable for practical deployment in artistic scenarios, and provides a reproducible engineering prototype for Chladni audio-visual linkage art creation.

\bibliographystyle{IEEEtran}
\bibliography{references}

\end{document}